\documentclass[conference]{IEEEtran}
\IEEEoverridecommandlockouts
\usepackage{cite}
\usepackage{amsmath,amssymb,amsfonts}
\usepackage{graphicx}
\usepackage{textcomp}
\usepackage[table]{xcolor}
\usepackage[hidelinks]{hyperref}
\usepackage{mdframed}
\usepackage{tabularray}
\usepackage{tikz}
\newcommand*\circled[1]{\tikz[baseline=(char.base)]{
            \node[shape=circle,draw,fill=black,text=white,inner sep=1pt] (char) {#1};}}
\usepackage{multirow}
\def\BibTeX{{\rm B\kern-.05em{\sc i\kern-.025em b}\kern-.08em
    T\kern-.1667em\lower.7ex\hbox{E}\kern-.125emX}}

\usepackage{algorithm,algpseudocode}
\newcommand{\algmargin}{\the\ALG@thistlm}
\makeatother
\algnewcommand{\parState}[1]{\State%
    \parbox[t]{\dimexpr\linewidth-\algmargin}{\strut\hangindent=\algorithmicindent \hangafter=1 #1\strut}}

\usepackage{wrapfig}
\usepackage{enumitem}
    
\begin{document}

\title{CMER: A Context-Aware Approach for Mining Ethical Concern-related App Reviews}

\author{
    \IEEEauthorblockN{Aakash Sorathiya, Gouri Ginde}
    \IEEEauthorblockA{\textit{Department of Electrical and Software Engineering} \\
        \textit{University of Calgary}\\
        Calgary, Canada \\
        \{aakash.sorathiya, gouri.ginde\}@ucalgary.ca}
}
\maketitle
\begin{abstract}
With the increasing proliferation of mobile applications in our daily lives, the concerns surrounding ethics have surged significantly. Users communicate their feedback in app reviews, frequently emphasizing ethical concerns, such as privacy and security. Incorporating these reviews has proved to be useful for many areas of software engineering (e.g., requirement engineering, testing, etc.). However, app reviews related to ethical concerns generally use domain-specific language and are typically overshadowed by more generic categories of user feedback, such as app reliability and usability. Thus, making automated extraction a challenging and time-consuming effort.

This study proposes CMER (A \underline{C}ontext-Aware Approach for \underline{M}ining \underline{E}thical Concern-related App \underline{R}eviews), a novel approach that combines Natural Language Inference (NLI) and a decoder-only (LLaMA-like) Large Language Model (LLM) to extract ethical concern-related app reviews at scale. In CMER, NLI provides domain-specific context awareness by using domain-specific hypotheses, and the Llama-like LLM eliminates the need for labeled data in the classification task. We evaluated the validity of CMER by mining privacy and security-related reviews (PSRs) from the dataset of more than 382K app reviews of mobile investment apps. First, we evaluated four NLI models and compared the results of domain-specific hypotheses with generic hypotheses. Next, we evaluated three LLMs for the classification task. Finally, we combined the best NLI and LLM models (CMER) and extracted 2,178 additional PSRs overlooked by the previous study using a keyword-based approach, thus demonstrating the effectiveness of CMER. These reviews can be further refined into actionable requirement artifacts.
\end{abstract}


\section{Introduction}
Mobile applications are designed with particular user goals in mind \cite{ebrahimi2022unsupervised}. For instance, the goal of investing apps is to enhance the financial capability of users \cite{bunnell2021development}. However, the competitive landscape of app development often leads to rapid production cycles that compromise the original user-focused goals \cite{ebrahimi2022unsupervised}, raising serious ethical issues such as privacy and security breaches \cite{chen2020empirical}. App marketplaces like the Apple App Store and Google Play allow users to rate and review apps, sharing their experiences regarding the app’s quality, performance, and concerns. Developers rely on these reviews to gather feedback and accordingly improve their apps \cite{biswas2024interpretable}.

Numerous studies have explored user perspectives on ethical concerns in mobile apps. Research by Besmer et al. \cite{besmer2020investigating} and Nema et al. \cite{nema2022analyzing} highlights user concerns about privacy and data security in mobile apps. The issues related to fairness and accessibility in mobile apps are further emphasized by the findings of Rezaei Nasab et al. \cite{rezaei2025fairness} and Reyes Arias et al. \cite{reyes2022accessibility}. However, these studies primarily rely on keyword- or regular expression-based sampling techniques, which may lack contextual information and restrict ethical concerns to specific keywords.  For instance, consider the following review:

\begin{quote}
    \textit{``Listen!! ok.. so why is there no phone number for customer service? why? if you are a great company and trustworthy why can't you get a phone line. only email?? c'mon sounds like you can take advantage if need be since nobody cant get threw only on email. if someone is missing money and you don't follow threw that can be scary esp if people invested \$1,000 in this. also, the glitch with updating id and social security stating it's a photocopy when it is a real deal. no good."}
\end{quote}

Here, the ``social security" phrase is mentioned in regards to a glitch in document uploads functionality; however, it was considered as a privacy-indicative keyword in the previous study \cite{ebrahimi2022unsupervised}. Consequently, the identification of ethical concerns-related reviews is significantly dependent on contextual interpretation; thus, merely conducting searches for related keywords within the review might not be an effective approach. Harkous et al. \cite{harkous2022hark} proposed the Natural Language Inference (NLI) method to address this. However, their approach relies on generic privacy hypotheses, overlooking the fact that users' ethical concerns are domain-dependent \cite{ebrahimi2022unsupervised}. For example, users of ridesharing apps may worry about constant location tracking, while those on financial apps may be concerned about sharing sensitive financial information. Moreover, all of these studies are based on the supervised classification of app reviews, which requires manual efforts and can be time-consuming.



To address these challenges, in this paper, we propose a novel hybrid approach: \textbf{CMER} (A \underline{C}ontext-Aware Approach for \underline{M}ining \underline{E}thical Concern-related App \underline{R}eviews), that combines NLI and a decoder-only Large Language Model (Llama-like LLM\footnote{In this paper, we refer to decoder-only LLMs, such as Llama, ChatGPT, and Claude, as Llama-like LLMs}) to mine ethical concern-related app reviews at scale. CMER utilizes NLI with domain-specific hypotheses (providing domain-specific context awareness) to extract potential concern-related app reviews. It then employs a Llama-like LLM with zero-shot learning (without any fine-tuning) to ultimately extract the actual concern-related app reviews from a set of potentially related reviews. We evaluate the effectiveness of CMER by mining privacy- and security-related reviews (PSRs) from mobile investment apps, guiding our research through three research questions (RQs).

\textit{RQ1. To what extent can domain-specific NLI accurately identify potential PSRs in investment apps?}
NLI assesses the relationship between a premise and a hypothesis to identify whether the premise entails, contradicts, or is neutral to the hypothesis \cite{fazelnia2024lessons}. NLI offers a more refined and linguistically consistent evaluation of text relationships, surpassing keyword-based methods by delivering context-aware results \cite{fazelnia2024lessons}. Therefore, we investigate to what extent we can use NLI with domain-specific hypotheses to identify the potential PSRs. This helps to filter out the large set of unrelated app reviews and focus on only potentially relevant app reviews. Additionally, we compare our domain-specific hypotheses with generic hypotheses provided by the previous study \cite{harkous2022hark} and show that domain-specific hypotheses yield better results due to the domain dependence of users' ethical concerns.

\textit{RQ2. To what extent can we leverage LLaMA-like LLMs to classify PSRs?}
NLI helps to extract potential PSRs from a large corpus of irrelevant app reviews however, further steps are required to extract actual PSRs from the set of potential reviews. Previous studies have employed supervised classification techniques, which required fine-tuning classification models on the labeled dataset. However, this technique requires manual efforts to create the labeled data, which can be labor-intensive and time-consuming. Therefore, we investigate to what extent we can leverage Llama-like LLMs with zero-shot learning, which do not require any fine-tuning on the labeled data, to classify PSRs.

\textit{RQ3. How effective is CMER compared to traditional keyword- and regex-based approaches in mining PSRs?}
After evaluating NLI and LLM individually, we select the best-performing models at both stages and compare our context-aware approach (CMER) with traditional keyword- and regex-based approaches. Our goal is to evaluate CMER for extracting PSRs that do not contain any predefined terms used in the keyword-based method by the previous study \cite{ebrahimi2022unsupervised}.

The main contributions of our study are:
\begin{itemize}[leftmargin=*]
    \item To our knowledge, CMER is the first hybrid approach utilizing domain-specific NLI and a Llama-like LLM to extract ethical concern-related app reviews. CMER demonstrated better results compared to traditional keyword-based and generic privacy hypotheses methods.
    \item We develop domain-specific hypotheses based on the finance domain-specific privacy and security risks taxonomy provided by Chen et al. \cite{chen2020empirical}.
    \item We demonstrated the efficacy of CMER by performing qualitative analysis on the dataset of more than 382K app reviews from mobile investment apps and extracted 2,178 additional PSRs that remained undetected by the previous study using a keyword-based method. These PSRs can be refined into actionable requirement artifacts to address users' privacy and security concerns.
    \item We open source our source code and dataset\footnote{\url{https://doi.org/10.5281/zenodo.15650122}} of 2,178 PSRs to support reproducibility and future research.
\end{itemize}

The rest of the paper is organized as follows. Section \ref{prem} explains the preliminaries, while Sections \ref{data} and \ref{method} detail the dataset and explain our research methodology, respectively. Section \ref{results} presents the results of our RQs, and Section \ref{discuss} discusses the implications of our findings. Section \ref{threats} identifies threats to validity, and Section \ref{rw} reviews related work. Finally, Section \ref{conclude} presents the conclusion and future directions.

\section{Background} \label{prem}
\textbf{Natural Language Inference (NLI):}
NLI pertains to the problem of ascertaining whether a natural language hypothesis can logically be derived from a specified premise \cite{maccartney2009extended}. An NLI model is required to evaluate whether a hypothesis is true (i.e., \textit{entailment}), false (i.e., \textit{contradiction}), or undetermined (i.e., \textit{neutral}) concerning a given premise. For instance, consider a premise stating, ``…their system is clearly not secure with your bank information…". A hypothesis asserting, ``The user is concerned about the security of their stored financial information" would be assigned an \textit{entailment} label. Conversely, a hypothesis claiming ``The user likes the secure data storage" would be designated a \textit{contradiction} label, and a hypothesis positing ``The app has a good interface" would be assigned a \textit{neutral} label.

Moreover, this methodology mitigates the dependency on specific keywords due to the extensive linguistic variability present in the premises associated with the hypotheses. For example, both of the following reviews receive an \textit{entailment} label for the hypothesis ``The user is concerned about unauthorized collection of sensitive financial information.":

\begin{itemize}
    \item ``I gave them my ssn twice, my real name...and my address twice. that worried me. then they wanted my bank info. then they wanted a picture of my bank statement..." (P(entailment)=0.99)
    \item ``It asks for way too much personal information for a telephone app!! i am not comfortable at all putting in my date of birth, complete address and, the final straw, my social security number!!! all this 8nfornation is going to someone (or group) i have absolutely no idea who they are!! do not trust it!" (P(entailment)=0.98)
\end{itemize}

Note that no review has any words in common with hypotheses, but both of them discuss the concern related to excessive financial data collection. Here, P(entailment) denotes the probability of the \textit{entailment} label and is referred to as \textit{entailment\_score}. We use these scores to filter out the potential reviews based on the defined heuristics.

\begin{algorithm*}
\caption{\textbf{RQ1}: NLI inference - Identifying the best NLI model and the best set of hypotheses with corresponding heuristics}\label{alg:cap}
\begin{algorithmic}[1]
\State \textbf{Input:} 31 generic hypotheses from \cite{harkous2022hark}, Heuristics \cite{harkous2022hark}, 17 newly defined domain-specific hypotheses (Table \ref{tab:dhypo}) and corresponding heuristics, and a labeled dataset of 3,519 app reviews from \cite{ebrahimi2022unsupervised}
\State \textbf{Output:} Best performing NLI model, Best of the two sets of hypotheses and corresponding heuristics, and Pseudo labeled corpus using best performing NLI model and best hypotheses
\State $generic\_hypotheses \gets $\text{31 hypotheses from \cite{harkous2022hark}}, $heuristics \gets \text{set of heuristics from \cite{harkous2022hark}}$
\State $NLI\_models \gets \text{[Roberta-large-mnli, Nli-roberta-base, DeBERTa-v3-base-mnli-fever-anli, T5-base]}$
\State $domain\_specific\_hypotheses \gets \text{[domain-specific hypotheses defined in Table \ref{tab:dhypo}]}$
\State $new\_heuristics \gets \text{[newly defined heuristics]}$, $dataset \gets \text{ground truth data from \cite{ebrahimi2022unsupervised}}$

\State $best\_NLI\_model = NLI\_models[0]$,  $best\_F1\_score = 0$
\For{$model \in NLI\_models$}
    \State $entailment\_scores = NLI\_Inference(model, generic\_hypotheses, dataset)$
    \State $nli\_annotated\_corpus = Apply\_Heuristics(entailment\_scores, heuristics)$
    \State $P, R, F1 = \text{\textit{Compute (Precision, Recall, and F1)}}$
    \If{$F1 > best\_F1\_score$}
        \State $\textbf{best\_NLI\_model} = model$
        \Comment {Determine the best performing NLI model on generic hypotheses and heuristics}
        \State $\textbf{best\_F1\_score} = F1$
    \EndIf
\EndFor

\State Next, use the best-performing NLI model on the domain-specific hypotheses
\State $entailement\_scores = NLI\_Inference(best\_NLI\_model, domain\_specific\_hypotheses, dataset)$
\State $nli\_annotated\_corpus = Apply\_Heuristics(entailment\_scores, new\_heuristics)$
\State $P, R, F1 = \text{\textit{Compute (Precision, Recall, and F1)}}$
\Comment{Metrics for best-performing NLI model with domain-specific hypotheses and new heuristics}
\State Next, determine which set of hypotheses is best performing by comparing F1 scores
\If {$F1 > best\_F1\_score$}
    \State $\textbf{best\_hypotheses} = domain\_specific\_hypotheses$
    \State $\textbf{best\_F1\_score} = F1$
\Else
    \State $\textbf{best\_hypotheses} = generic\_hypotheses$
\EndIf
\State $\textbf{pseudo\_labeled\_corpus} = labels(dataset, best\_NLI\_model, best\_hypotheses)$
\Comment{Corpus containing `maybe-psr' and `maybe-not-psr' labels}
\end{algorithmic}
\end{algorithm*}

\section{Dataset} \label{data}
\begin{table}[h]
    \centering
    \renewcommand{\arraystretch}{1.1}
    \caption{Statistics of the dataset used from Ebrahimi et al. \cite{ebrahimi2022unsupervised}. We utilize 1-2 star-rated reviews in our experiments, as Ebrahimi et al. have shown that users' privacy and security concerns are related to low star ratings.}
    \label{tab:stat}
    \begin{tabular}{l|l}
        \textbf{Number of apps} & 8 \\
        \hline
        \textbf{App domain} & Finance (Investment) \\
        \hline
        \textbf{Total reviews} & 696,073 \\
        \hline
        \textbf{Total 1-2 star rated reviews} & 385,951 \\
        \hline
        \textbf{Labeled PSRs (1-2 star rated)} & 1,058 \\
        \hline
        \textbf{Labeled non-PSRs (1-2 star rated)} & 2,461 \\
        \hline
        \textbf{Average number of words per review} & 27 \\
        \hline
        \textbf{Time range} & 2009-04-24 to 2021-06-15
    \end{tabular}
\end{table}

To assess the validity of CMER we utilized the dataset of app reviews from mobile investing apps provided by Ebrahimi et al. \cite{ebrahimi2022unsupervised}. Table \ref{tab:stat} shows statistical information about the dataset. It contains a total of 696,073 app reviews from the most widely used investing applications available on the Google Play Store and the Apple App Store. It includes 385,951 1-2 star-rated reviews, from which 3,519 reviews are manually validated and labeled with 0 (2,461 non-PSRs) and 1 (1,058 PSRs) labels. The data collection and labeling process are presented in detail in the original study \cite{ebrahimi2022unsupervised}. In this study, we utilized the labeled dataset of 3,519 reviews to answer RQ1 and RQ2. For RQ3, we used unlabeled 382,432 1-2 star-rated reviews to extract additional PSRs using CMER that did not contain any predefined set of keywords.

\section{Methodology} \label{method}
\begin{figure*}[h]
    \centering
    \includegraphics[scale=0.26]{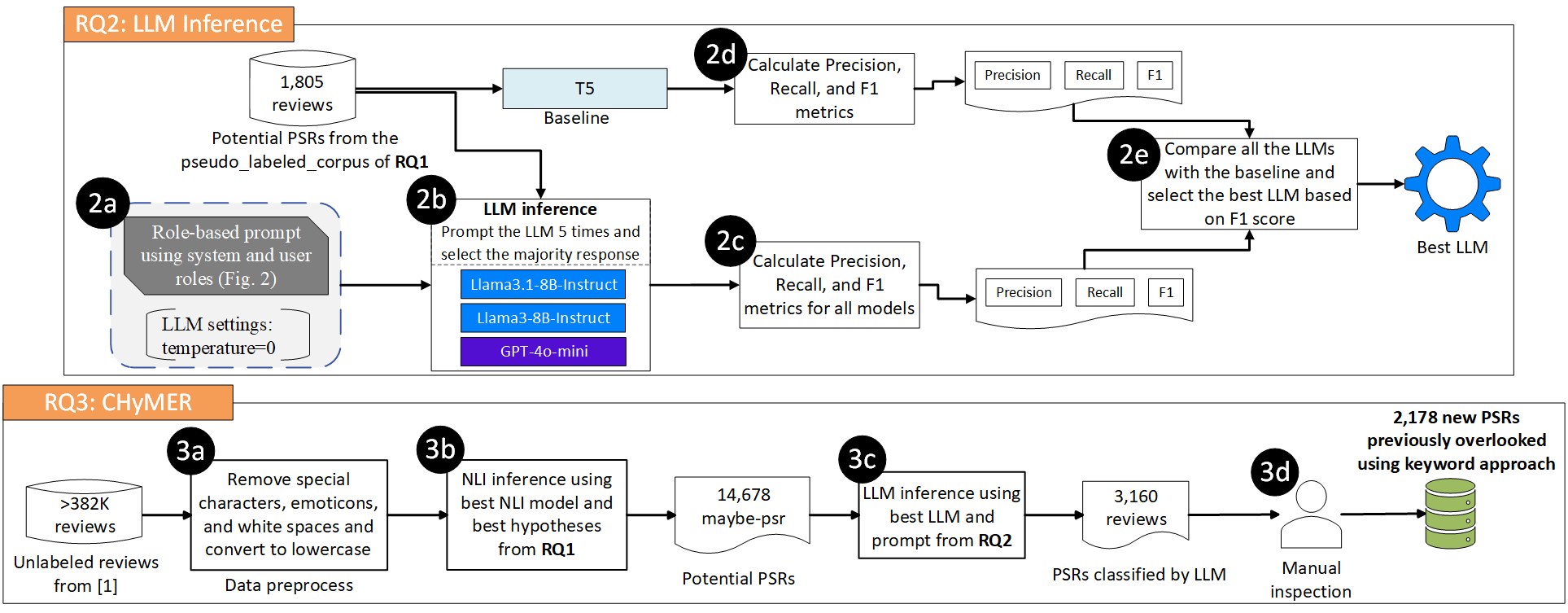}
    \caption{Overview of our methodology for LLM inference (RQ2) and extracting concern-related reviews using CMER (RQ3).}
    \label{fig:study}
\end{figure*}
Algorithm \ref{alg:cap} and Figure \ref{fig:study} provide an overview of our research methodology. Our approach consists of three parts: (1) NLI-inference: we identify the best NLI model and the best set of hypotheses to extract potential PSRs; (Algorithm \ref{alg:cap}) (2) LLM-inference: we then compare the performance of various LLaMA-like LLMs to identify the best-performing LLM for the classification of PSRs from a set of potential PSRs (Figure \ref{fig:study}); (3) Finally, we combine NLI from RQ1 (NLI-inference) and LLM from RQ2 (LLM inference) to extract additional PSRs (Figure \ref{fig:study}).
We detailed the methods employed in each component below.

\underline{\textbf{1) NLI Inference:}}
Algorithm \ref{alg:cap} describes the proposed NLI inference process. Using 31 generic hypotheses and the corresponding heuristics from Harkous et al. \cite{harkous2022hark}, and the labeled dataset of 3,519 app reviews \textbf{(lines 1-7)}, we first determined the best NLI model \textbf{(lines 8-16)} of the four chosen NLI models, namely: Roberta-large-mnli, Nli-roberta-base, DeBERTa-v3-base-mnli-fever-anli, and T5-base using Precision (P), Recall (R), and F1-score as measures.  We performed 3,519 (number of app reviews) * 31 (generic hypotheses) * 4 (number of models) = 436,356 inference operations at this stage.

Next, to determine the best hypotheses, we compared the performance of the generic hypotheses (baseline) with the newly defined domain-specific hypotheses and respective corresponding heuristics \textbf{(lines 17-27)}. We performed 3,519 (number of app reviews) * 17 (domain-specific hypotheses) = 59,823 inference operations. In the end (\textbf{line 28}), we use the best NLI model and best hypotheses with their corresponding heuristics to create a pseudo-labeled corpus containing `maybe-psr' and `maybe-not-psr' labels. This pseudo-labeled corpus is further used for the evaluation of LLMs in RQ2.

\textit{Determining domain-specific hypotheses and corresponding heuristics:}
To derive the finance domain-specific hypotheses, we utilized the taxonomy of privacy and security risks of Android finance applications provided by Chen et al. \cite{chen2020empirical}. For each risk category in the taxonomy, we manually defined one or more hypotheses representing the risk in the form of a natural language sentence, which can be used to find the app reviews related with an \textit{entailment} label. For example, for the ``Input Harvest" risk category, we defined five hypotheses, such as ``The app collects more financial data than necessary" and ``The user is concerned about how the app harvests their financial data".
In total, we defined 17 domain-specific hypotheses as shown in Table \ref{tab:dhypo}.

Further, to select the potential PSRs based on the \textit{entailment\_scores}, we defined the following heuristics based on the experiments performed on a different dataset\footnote{This study is part of a broader study, where part of it is submitted in another track of the RE conference, in which we performed the experiments with different sets of heuristics and the best one is selected for this study.}. Here, N\textsubscript{E}(i, t) is the number of hypotheses receiving an \textit{entailment\_scores} above a threshold t for review i.

\begin{itemize}
    \item A review i is labeled as \textit{maybe-psr} if \textit{N\textsubscript{E}(i, 0.85)$>$=1} or \textit{N\textsubscript{E}(i, 0.75)$>$=3} or \textit{N\textsubscript{E}(i, 0.7)$>$=5}.
    \item The rest of the reviews are labeled as \textit{maybe-not-psr}.
\end{itemize}

In our experiments, this set of heuristics demonstrated the best performance by yielding the lowest number of false positives (FP) (0-labeled reviews annotated as `maybe-psr') and false negatives (FN) (1-labeled reviews annotated as `maybe-not-psr').

\begin{table}[h]
    \centering
    \caption{Finance domain privacy and security risks taxonomy \cite{chen2020empirical} and associated hypotheses.}
    \label{tab:dhypo}
    \begin{tabular}{p{1.7cm}|p{6.5cm}}
        \hline
        \textbf{Risk Category} & \textbf{Hypotheses} \\
        \hline 
        \multirow{1}{1.5cm}{Input Harvest} & 1. The user is concerned about how the app harvests their financial data. \\
        & 2. The user is concerned about unauthorized collection of sensitive financial information. \\
        & 3. The app requires excessive permissions to access financial data. \\
        & 4. The app collects financial data without adequate transparency. \\
        & 5. The app collects more financial data than necessary. \\
        \multirow{1}{1.5cm}{Sensitive Data Storage} & 6. Financial data is retained for longer than necessary. \\
        & 7. The user is concerned about the security of their stored financial information. \\
        & 8. The app stores sensitive financial data without proper encryption. \\
        & 9. The user is concerned about the processing and storage of financial data against privacy regulations or policies. \\
        & 10. The user is concerned that their financial data is stolen due to hacking. \\
        \multirow{1}{1.5cm}{Sensitive Data Transmission} & 11. The user is concerned about the interception of their financial transactions. \\
        & 12. Financial data is shared with third parties during transmission without consent. \\
        & 13. Sensitive financial data is shared with third parties for marketing or profit. \\
        & 14. Financial information is accessible to internal firm advisors without consent. \\
        \multirow{1}{1.6cm}{Communication Infrastructure} & 15. Sensitive financial details are shared via insecure channels. \\
        & 16. Unauthorized bank transfers are performed. \\
        & 17. User device communication patterns reveal private financial information. \\
        \hline
    \end{tabular}
\end{table}

\textit{NLI Models:} We performed inference with four different NLI models. These models were chosen due to their state-of-the-art NLI performance and easy availability on the HuggingFace platform \cite{wolf2019huggingface}. Additionally, these models are fine-tuned and pre-trained for the NLI task using state-of-the-art NLI datasets, namely: Multi-Genre Natural Language Inference (MNLI) \cite{williams2017broad} (433k sentence pairs), Adversarial Natural Language Inference (ANLI) \cite{nie2019adversarial} (169k sentence pairs), Stanford Natural Language Inference (SNLI) \cite{bowman2015large} (570k sentence pairs), Question Answer NLI (QNLI) \cite{rajpurkar2016squad} (116k sentence pairs), Fact Extraction and VERification NLI (FeverNLI) \cite{thorne2019fever2} (185k sentence pairs).

\noindent The four chosen NLI models are as follows:
\begin{itemize}[leftmargin=*]
    \item Roberta-large-mnli: is the RoBERTa large model \cite{liu2019roberta} fine-tuned on the MNLI dataset. The model is pre-trained on English language text using a masked language modeling (MLM) objective.
    \item Nli-roberta-base: is the RoBERTa base model \cite{liu2019roberta} fine-tuned on the MNLI and SNLI datasets using Sentence Transformers Cross-Encoder class \cite{reimers-2019-sentence-bert}.
    \item DeBERTa-v3-base-mnli-fever-anli: is the DeBERTa-v3 base model \cite{he2021deberta} fine-tuned on the MNLI, FeverNLI, and ANLI datasets.
    \item T5-base: is the vanilla T5 model \cite{2020t5} readily fine-tuned on the MNLI and QNLI datasets.
\end{itemize}

\noindent To implement these models, we used the transformers library from HuggingFace \cite{wolf2019huggingface} with their respective tokenizers. 

\textit{Generic hypotheses and corresponding heuristics (baseline for RQ1):}
We assessed the validity of our finance domain-specific hypotheses by comparing them with the set of 31 generic hypotheses (not related to any particular domain) and corresponding heuristics (both available in our supplementary material included in our replication package) provided by Harkous et al. \cite{harkous2022hark}. 
These hypotheses were derived from Solove's \cite{solove2005taxonomy} taxonomy of privacy violations and the taxonomy of privacy-enhancing technologies proposed by Wang and Kobsa \cite{wang2009privacy}. 

\underline{\textbf{2) LLM Inference:}}
Figure \ref{fig:study} outlines the proposed LLM inference process. We utilized the potential PSRs (annotated as `maybe-psr') from the pseudo-labeled corpus created in the NLI inference component.
Initially, we formulated the prompt and configured the LLM parameter as illustrated in step \circled{2a}. Subsequently, we selected four distinct LLaMA-like LLMs, executed the inference procedure (step \circled{2b}), and computed the P, R, and F1-score metrics for the obtained results (step \circled{2c}). Finally, we calculated these metrics for the baseline T5 model (step \circled{2d}) and compared the performance of LLaMA-like LLMs against the baseline model to identify the best-performing LLM (step \circled{2e}).

\textit{Choice of LLaMA-like LLMs:}
To make our study replicable and more accessible, we chose two open-source LLMs that have shown state-of-the-art performance for various NLP tasks \cite{patil2024review} and are readily available through the transformers library of HuggingFace \cite{wolf2019huggingface}. We also selected one closed-source LLM to perform a comparative analysis with open-source LLMs.

\begin{itemize}[leftmargin=*]
    \item meta-llama/Llama-3.1-8B-Instruct \cite{dubey2024llama}: is the LLaMA 3.1 instruction-tuned text-only auto-regressive language model that uses an optimized transformer architecture. The tuned versions use supervised fine-tuning (SFT) and reinforcement learning with human feedback (RLHF) to align with human preferences for helpfulness and safety.
    \item meta-llama/Meta-Llama-3-8B-Instruct \cite{dubey2024llama}: is the LLaMA 3 instruction-tuned text-only auto-regressive language model.
    \item GPT-4o-mini \cite{ye2023comprehensive}: Several GPT LLMs have recently been proposed, with a prevalence of studies using the GPT3.5-turbo \cite{ouyang2023llm}. However, GPT-4o-mini (the model that powers ChatGPT at the time of writing this) is relatively affordable, capable, and fast, and is an improved version of GPT3.5-turbo \cite{ye2023comprehensive}. Therefore, we decided to use GPT-4o-mini \cite{hurst2024gpt}.
\end{itemize}

For open-source models, we specifically used the Instruct version because it is specifically fine-tuned using the Reinforcement Learning for Human Feedback (RLHF) technique, to follow the user instructions provided in the prompt \cite{ouyang2022training}.

\textit{Prompt design and parameter configuration for LLMs:}
\begin{figure}[b]
    \centering\includegraphics[scale=.16]{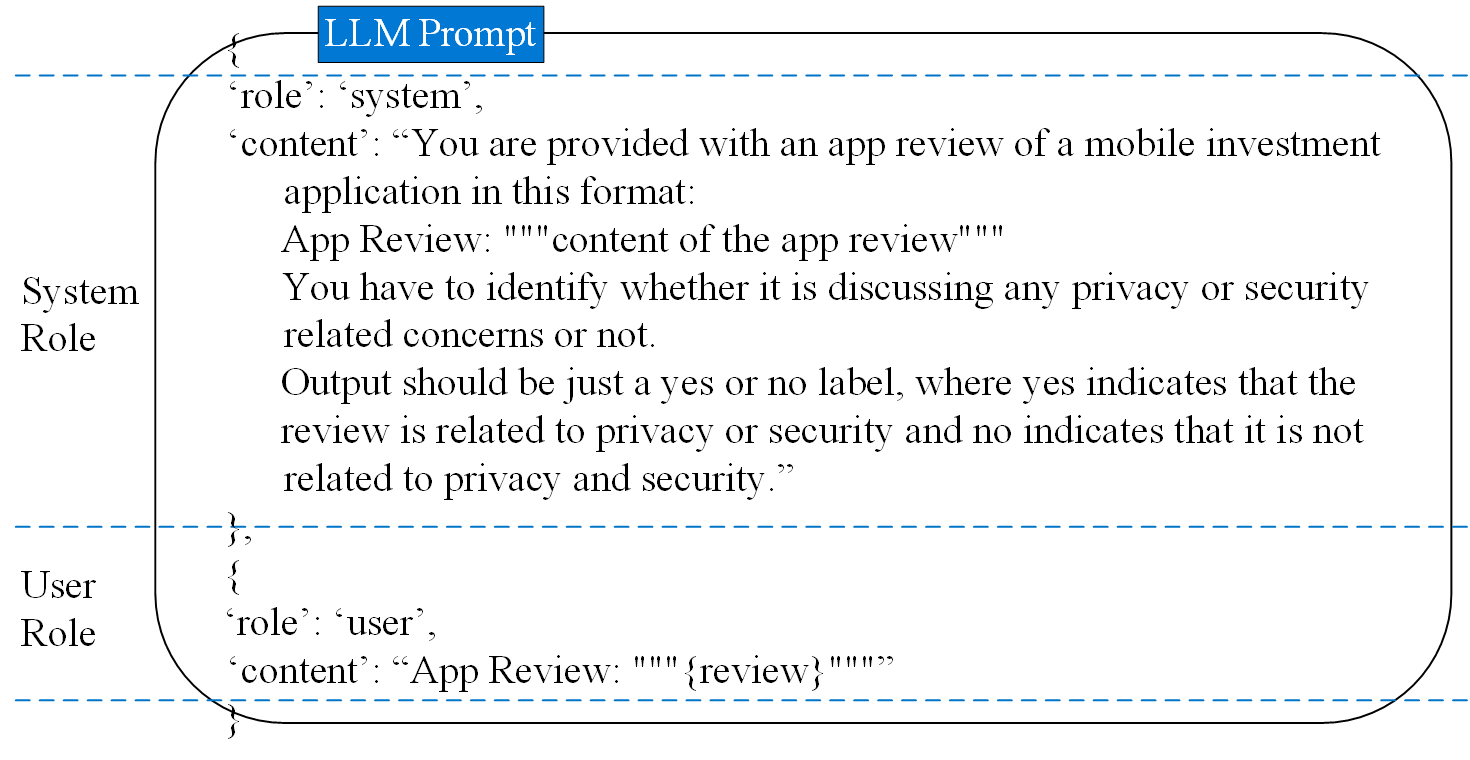}
    \vspace{-2mm}
    \caption{Role-based prompt designed following the guidelines from \cite{chen2023unleashing}.}
    \label{fig:prompt}
\end{figure}
In line with previous work \cite{touvron2023llama, zhang2023revisiting}, we designed a prompt for LLMs to experiment with zero-shot learning following the guidelines provided by \cite{chen2023unleashing}. We followed the role-based prompt technique by defining clear and precise instructions for each role. This technique involves giving the model a specific role, such as a helpful assistant or an expert \cite{chen2023unleashing}. It can be particularly effective in guiding model responses and ensuring that they align with the desired output \cite{chen2023unleashing}. In our design, we defined two roles, system and user, for the model. In the system role, we first defined the format of the input provided to the model, comprising an app review as \textit{App Review: ``````content of the app review"""}. Next, we described the classification task to instruct the model on what to do, and in the end, we provided the desired format for the model output, where we specifically asked the model to return just the \textit{yes}/\textit{no} labels indicating whether the input app review is related to privacy/security or not. In the user role, we provided the model with the app review in the desired format. Figure \ref{fig:prompt} shows the detailed prompt structure.

The LLM setting, specifically the model temperature, plays a crucial role in generating responses \cite{chen2023unleashing}. It controls the randomness of the generated output, i.e., a lower temperature leads to more deterministic outputs \cite{chen2023unleashing}. Thus, we set the temperature value to 0 to obtain a deterministic response and minimize the likelihood of generating incorrect or unpredictable responses. Furthermore, we prompted the model five times and selected the majority response to overcome the inherent variability in model responses and increase the chances of obtaining a more deterministic output \cite{chen2023unleashing}.

\textit{Text-to-Text Transfer Transformer (T5) (baseline for RQ2)}:
There are several supervised classification models proposed by previous studies to classify app reviews as PSRs, including SVM \cite{nguyen2019short}, BERT \cite{nema2022analyzing}, and T5 \cite{harkous2022hark}. However, T5 has shown state-of-the-art performance compared to other supervised classification models in the previous study \cite{harkous2022hark}, therefore, we chose T5 as our baseline to compare the results of Llama-like models, which do not require any fine-tuning on the labeled dataset. T5 is an encoder-decoder model with a unified architecture based on casting problems into the text-to-text paradigm and training a model on the text generation objective. We cast the task of classifying app reviews into text-to-text generation and fine-tuned the T5 base model, reproducing the one used in Harkous et al. \cite{harkous2022hark}.

\underline{\textbf{3) CMER}:} 
Figure \ref{fig:study} illustrates the complete flow of the CMER (combining the best NLI model with domain-specific hypotheses and the best LLM). We performed the qualitative evaluation of CMER by comparing it with the traditional keyword-based method \cite{ebrahimi2022unsupervised} (baseline for RQ3) and tried to extract additional PSRs that did not contain any predefined keywords. We utilized the set of 382,432 unlabeled app reviews from Ebrahimi et al. \cite{ebrahimi2022unsupervised} and preprocessed all the reviews to remove special characters, emoticons, and white spaces, and converted them to lowercase (step \circled{3a}). Next, we performed the NLI inference using the best NLI model and the best set of hypotheses from RQ1 to filter out the non-relevant reviews and get a set of potential PSRs (maybe-psr) (step \circled{3b}). Then, we performed LLM inference using the best LLM from RQ2 to get the actual PSRs from the set of potential reviews (step \circled{3c}). Finally, in the end (step \circled{3d}), we performed the manual inspection (detailed below) to create a ground truth dataset of PSRs that remained previously unidentified.

\textit{Manual inspection setup:}
The manual inspection was performed to validate all the reviews extracted using CMER and create a ground truth dataset of PSRs supporting future research in the domain. Four annotators, including the first author and 3 graduate students from our research lab, conducted this task. The first author analyzed all the reviews, while the others inspected one-third of the sample so that each review was inspected at least twice. To avoid exhaustion, we performed this process in a 10-day period. To ensure the understanding of the task and the definitions of PSR and non-PSR labels, we created the labeling instructions (available in our replication package) and we based our analysis on the taxonomy provided by Chen et al. \cite{chen2020empirical} to label the reviews. After the manual inspection, we cross-checked the findings of the manual classification. For every disagreement, a third annotator was requested to break the tie. In total, 299 reviews had to be further analyzed by another annotator. To determine the extent to which the annotators agreed upon the classifications, we used Cohen’s $kappa$ coefficient \cite{cohen1968weighted}. We acquired a degree of agreement of 0.78, which shows substantial agreement and indicates high reliability in the annotation process \cite{viera2005understanding}.

\underline{\textbf{4) Evaluation measures:}}
To evaluate the performance of NLI and LLM inference operations, we employed the measures of P, R, and F1-score. The F1-score (\(F1 = \frac{2*P*R}{P+R}\)) corresponds to the harmonic mean of P (\(P = \frac{TP}{TP+FP}\)) and R (\(R = \frac{TP}{TP+FN}\)), where P is the number of correct predictions out of all the input sample and R is number of positive predictions observed in the actual class.

In case of NLI evaluation, True Positives (TP) refers to the number of 1-labeled reviews annotated with a `maybe-psr' label, True Negatives (TN) refers to the number of 0-labeled reviews annotated with `maybe-not-psr' label, FP refers to the number of 0-labeled reviews annotated with `maybe-psr' label and FN refers to the number of 1-labeled reviews annotated with the `maybe-not-psr' labels.

In case of LLM evaluation, TP refers to the number of 1-labeled reviews classified with the `yes' label, TN refers to the number of 0-labeled reviews classified with the `no' label, FP refers to the number of 0-labeled reviews classified with the `yes' label and FN refers to the number of 1-labeled reviews classified with the `no' label.

\textbf{\underline{5) Computational Resources}}
The experiments were conducted on an NVIDIA GeForce RTX 4090 GPU of 40 GB of RAM and a 24-core CPU setup. We implemented our models using Python 3.12 with CUDA version 11.8 and HuggingFace Transformers version 4.44.1. We used NumPy and Pandas for linear algebra operations.

\section{Results} \label{results}
In this section, we present the results of our RQs.

\textit{RQ1 - To what extent can domain-specific NLI accurately identify potential PSRs in investment apps? }
To answer this question, first, we evaluated four NLI models using the baseline generic privacy hypotheses and selected the best NLI model based on the highest F1-score. Table \ref{tab:rq1} shows the inference results for the generic hypotheses. DeBERTa-v3-base-mnli-fever-anli performed best with the highest F1-score of 0.59. It can be observed that all the models achieved low precision as NLI identifies a high number of FPs. DeBERTa-v3-base-mnli-fever-anli model with generic hypotheses annotated 1,855/3,519 reviews as `maybe-psr', with 207 1-labeled reviews as `maybe-not-psr' and 1,004 0-labeled reviews as `maybe-psr'. Additionally, it can be observed that the Roberta-large-mnli and Nli-robert-base models performed similarly, whereas the T5-base model achieved the lowest performance.

Next, we compared the inference results of the domain-specific hypotheses with generic hypotheses using the best-performing DeBERTa-v3-base-mnli-fever-anli model. Table \ref{tab:rq1_2} shows the findings indicating that domain-specific hypotheses yielded better results compared to generic hypotheses. We achieved an F1-score of 0.65 with domain-specific hypotheses compared to a 0.56 F1-score with generic hypotheses. This improvement is promising in terms of FP and FN as highlighted in Table \ref{tab:rq1_2}. In terms of FP, we identified only 878 FPs with domain-specific hypotheses, whereas this count was comparatively higher (1,004) for generic hypotheses. For FN, we received a lower count of 131 FNs with domain-specific hypotheses compared to 207 FNs with generic hypotheses.

\begin{mdframed}[backgroundcolor=white]
Summary of RQ1: The DeBERTa-v3-base-mnli-fever-anli NLI model with domain-specific hypotheses and corresponding heuristics performed best in extracting potential PSRs. It achieved an F1-score of 0.65, outperforming baseline generic hypotheses.
\end{mdframed}

\begin{table}[t]
    \centering
    \renewcommand{\arraystretch}{1.2}
    \caption{Results of NLI inference using the baseline generic hypotheses and corresponding heuristics.}
    \label{tab:rq1}
    \begin{tabular}{p{4.3cm}|p{.5cm}p{.5cm}p{.5cm}}
        \textbf{Model} & \textbf{P} & \textbf{R} & \textbf{F1} \\ \hline
        Roberta-large-mnli & 0.42 & 0.84 & 0.56 \\
        \textbf{DeBERTa-v3-base-mnli-fever-anli} & 0.46 & 0.80 & \textbf{0.59} \\
        T5-base & 0.39 & 0.49 & 0.43 \\
        Nli-roberta-base & 0.44 & 0.71 & 0.54 \\
    \end{tabular}
\end{table}
\begin{table}[t]
    \centering
    \renewcommand{\arraystretch}{1.2}
    \caption{Comparison of NLI inference using domain-specific hypotheses and baseline generic hypotheses with the best NLI model.}
    \label{tab:rq1_2}
    \begin{tabular}{p{2.3cm}|p{.5cm}p{.5cm}|p{.5cm}p{.5cm}p{.5cm}}
        \textbf{Hypotheses} & \textbf{FP} & \textbf{FN} & \textbf{P} & \textbf{R} & \textbf{F1} \\ \hline
        Generic (Baseline) & 1,004 & 207 & 0.42 & 0.84 & 0.56 \\
        \rowcolor{green!15} Domain-specific & 878 & 131 & 0.51 & 0.88 & 0.65
    \end{tabular}
\end{table}

\textit{RQ2 - To what extent can we leverage LLaMA-like LLMs to classify PSRs? }
To answer this question, we used 1,805 `maybe-psr' reviews from the pseudo-labeled corpus of the DeBERTa-v3-base-mnli-fever-anli model with domain-specific hypotheses (best performing ones from RQ1). Table \ref{tab:rq2} shows the P, R, and F1-score metrics of all the LLMs and the baseline T5 model, along with the number of TP, TN, FP, and FN for all the models. Our findings highlight that Llama-3.1-8B-Instruct outperformed the baseline T5 model (F1=0.80) in the zero-shot setting (without any fine-tuning), achieving an F1-score of 0.85. Furthermore, GPT-4o-mini outperformed T5 with an F1 score of 0.82, while Llama-3-8B-Instruct showed comparable performance with an F1 score of 0.80.

\begin{mdframed}[backgroundcolor=white]
Summary of RQ2: Llama3.1-8B-Instruct achieved the best performance for the classification of PSRs without any fine-tuning on the labeled dataset, with an F1-score of 0.85 and surpassing the fine-tuned baseline T5 model.
\end{mdframed}

\begin{table}[t]
    \centering
    \caption{LLM inference results on the dataset of 1,805 \textit{maybe-psr} reviews and their comparison with the baseline T5 model.}
    \label{tab:rq2}
    \renewcommand{\arraystretch}{1.1}
    \begin{tabular}{p{2.8cm}|p{.4cm}p{.4cm}p{.4cm}p{.4cm}|p{.4cm}p{.4cm}p{.4cm}}
        \textbf{Model} & \textbf{TP} & \textbf{TN} & \textbf{FP} & \textbf{FN} & \textbf{P} & \textbf{R} & \textbf{F1} \\
        \hline
        T5 (baseline) & 862 & 532 & 346 & 65 & 0.71 & 0.92 & 0.80 \\
        \rowcolor{green!15} Llama-3.1-8B-Instruct & 896 & 581 & 297 & 31 & 0.75 & 0.96 & \textbf{0.85} \\
        GPT-4o-mini & 878 & 602 & 318 & 44 & 0.73 & 0.95 & 0.82 \\
        Llama-3-8B-Instruct & 853 & 527 & 351 & 74 & 0.70 & 0.92 & 0.80 \\
    \end{tabular}
\end{table}

\textit{RQ3 - How effective is CMER compared to traditional keyword- and regex-based approaches in mining PSRs? }
To answer this question, we utilized the unlabeled dataset of 382,432 reviews and extracted additional PSRs using CMER. We used the DeBERTa-v3-base-mnli-fever-anli model (best-performing NLI model) with the domain-specific hypotheses and corresponding heuristics from RQ1 and the Llama3.1-8B-Instruct LLM (best-performing LLM) with the prompt from RQ2. After data preprocessing, we performed NLI inference and identified 14,678 `maybe-psr’ reviews. These reviews were then used in LLM inference operation, and 3,160 reviews were further labeled as `yes' by the LLM, indicating privacy and security-related reviews. After this, we performed the manual inspection and created a dataset of 2,178 PSRs that were not extracted by the previous study \cite{ebrahimi2022unsupervised} using keyword-based filtering. We show a few examples of additional PSRs in Section \ref{discuss} and make the whole dataset publicly available in our replication package.

\begin{mdframed}[backgroundcolor=white]
Summary of RQ3: CMER extracted 2,178 additional PSRs without any fine-tuning from a dataset of more than 382K app reviews.
\end{mdframed}

\section{Discussion} \label{discuss}
In this section, we summarize the findings and discuss the implications of our study.

\textbf{Discussing RQs.} The results of RQ1 show that the DeBERTa-V3 model fine-tuned on MNLI, FeverNLI, and ANLI datasets outperformed other NLI models, which can be attributed to its large NLI fine-tuning dataset containing 787K sentence pairs and its superior performance over the RoBERTa model in most natural language understanding tasks \cite{he2021deberta}. Further, the results of RQ2 show that the Llama3.1-8B model of the instruct version outperformed the baseline T5 model and other open-source (Llama3-8B-Instruct) and closed-source (GPT-4o-mini) LLMs, which can be attributed to its additional fine-tuning using the RLHF technique \cite{ouyang2022training} to follow user instructions. Based on these findings and in line with the previous study \cite{hanke2024open}, we suggest to use open-source LLMs to achieve privacy-preserving LLMs that yield relatively better performance compared to closed-source LLMs.

Our findings from RQ1 indicate that domain-specific hypotheses produce superior results compared to generic hypotheses, attributable to the domain-specific nature of users' ethical concerns. Furthermore, it is observed that all NLI models exhibited suboptimal precision values. However, from the findings of RQ3, we highlight the importance of using NLI to effectively filter the substantial amount of irrelevant data, as it successfully identified only 14,678 potential reviews from a large dataset of more than 382K reviews. App reviews encompass a significant proportion of irrelevant content, with only 0.5\% related to privacy and security \cite{mukherjee2020empirical}. Therefore, it is impractical to input all irrelevant information into our LLM classifier, as this would impede its overall performance.

Our RQ2 results highlight that the Llama3.1 model outperformed the supervised baseline T5 model in the classification of PSRs without any fine-tuning on the labeled dataset. Further, the RQ3 results highlight that CMER extracted only 3,160 relevant reviews from the dataset of more than 382K reviews for further manual inspection. These findings show the potential of using CMER to reduce the manual efforts associated with data labeling. Furthermore, based on these findings and in line with previous work \cite{zhang2023revisiting, zhu2023can}, we recommend integrating CMER into an annotation team to reduce manual efforts of human annotation \cite{wang2021want} or to assist with data augmentation \cite{moller2023parrot}. 

\begin{table*}[h]
    \renewcommand{\arraystretch}{1.2}
    \centering
    \caption{Examples of PSRs extracted using CMER.}
    \label{tab:rr}
    \begin{tabular}{lp{1.5cm}p{13.5cm}}
         \textbf{App} & \textbf{Date} & \textbf{Privacy Review}  \\
         \hline
         ETrade & 07/04/2013 & It records audio and video without your confirmation. Seems more like a corporate espionage or insider trading spy tool then a trading app with this type of capability.\\
         Fidelity & 08/10/2020 & Asks for void checks and so much more to access my bank account.\\
         Personal Capital & 05/02/2014 & They knew; 1) my total asset worth 2) which institutions (banks and brokerages) I use and 3) asset allocations of my portfolio. Not feeling too secure.\\
         Robinhood & 01/09/2020 & I had 5 different transactions totaling \$10,800 trying to be removed from my account by Robinhood. Currently trying to work on figuring this out, but their system is clearly not secure with your bank information.\\
         Schwab & 14/12/2017 & As soon as you download the app and enter your info they hit your credit. No where in the legal disclosure or terms and condition does it say anything about your FICO scores, credit checks, etc.\\
         Stash & 01/04/2021 & Someone hacked into my account, changed my email and password, and transferred my available funds to their account.\\
         TD Ameritrade & 16/02/2018 & The App sometimes (not always) does NOT automatically log you out after even many hours of non-activity. Please fix this major security risk!  Please provide an option to automatically log out after, say, 10 minutes of inactivity.\\
         \hline
    \end{tabular}
\end{table*}

\textbf{New ground truth dataset created using CMER.} Using CMER, we extracted 2,178 additional PSRs that did not contain any predefined keywords. 
We present a few reviews in Table \ref{tab:rr} and make the whole dataset publicly available. All these reviews discuss users' privacy and security concerns and are specifically related to the finance domain, including concerns related to sensitive financial information, such as confidential banking data and personal asset portfolios.
Furthermore, we conducted an automated analysis using wordcloud generation and bigram extraction methodologies, which revealed significant privacy and security concerns associated with account hacking, selling users' financial data, and stealing their money. 
This preliminary analysis shows the potential of CMER in extracting privacy- and security-related reviews, which can be refined into actionable requirement artifacts to address users' concerns. Our replication package includes a detailed analysis of these concerns.


\textbf{Abstract tool for app developers and adaptation of CMER to other domains.} CMER is an abstract tool comprising two primary components: NLI with domain-specific hypotheses and a decoder-only LLM. This tool helps app developers analyze user feedback and address their concerns. Furthermore, we highlight that CMER is adaptable to other domains, such as the sharing economy, health and fitness, or gaming, to extract reviews related to ethical concerns, such as safety, transparency, or accountability. Moreover, it can be adapted to analyze feedback from social media platforms such as Reddit \cite{li2022narratives}. This adaptability is achieved by defining hypotheses related to any ethical concern within any app domain and incorporating them into CMER, showcasing its reusability and adaptability across domains.

\section{Threats to Validity} \label{threats}

\textbf{Construct threats}:
Developing a dataset is a tedious job and also subject to reader bias. Therefore, we employed a methodological approach for manual inspection, including four annotators to mitigate the risk of individual bias, and conducted this process in 10 days timeframe to avoid fatigue. Further, we evaluated four NLI models and four LLMs; however, we acknowledge that applying other models to our dataset may lead to different results. Additionally, P, R, and F1 metrics used in this study are widely used in the SE domain and recommended for such classification tasks.

\textbf{Internal threats}: 
The process of defining domain-specific hypotheses and corresponding heuristics, along with LLM prompt design and parameter setting, may introduce internal validity threats. We used the technique suggested by previous studies to mitigate these threats. Similar to \cite{harkous2022hark}, we defined domain-specific hypotheses based on the widely used privacy and security risk taxonomy for finance apps provided by Chen et al. \cite{chen2020empirical}. Further, we followed the approach of \cite{harkous2022hark, duvsek2020evaluating} to define our corresponding heuristics; however, we recognize that altering these could affect the results. To design the prompt for LLMs, we followed the guidelines provided by \cite{chen2023unleashing}. Additionally, we experimented with providing the domain-specific hypotheses in the prompt; however, it did not improve the results. Furthermore, we set the temperature value to 0 for LLMs for all experimental conditions. However, future studies might consider investigating the impact of different values of temperature for classification tasks.

\textbf{External threats}: 
Future studies can explore few-shot learning to reduce the risk of model bias and hallucination. Future replications with larger and more varied benchmark datasets, including app reviews from different domains and platforms, are also required. These will enable evaluating the generalizability of our findings, thus assessing how consistent the behavior of CMER is across different models and datasets.

\section{Related Work} \label{rw}

\begin{table*}[t]
    \centering
    \renewcommand{\arraystretch}{1.2}
    \caption{Details of related works in comparison to CMER.}
    \label{tab:rw}
    \begin{tabular}{p{2.9cm}p{1.8cm}p{2.1cm}p{2.5cm}p{2.5cm}p{2.9cm}}
        \textbf{Study} & \textbf{App Domain} & \textbf{Concern Domain} & \textbf{Data Sampling} & \textbf{Classifier Model} & \textbf{Fine-tuning Required?} \\
        \hline
        Besmer et al. \cite{besmer2020investigating} & General & Privacy/Security & Keywords & Logistic regression & Yes \\
        Mukherjee et al. \cite{mukherjee2020empirical} & General & Privacy/Security & Keywords & SVM & Yes \\
        Nguyen et al. \cite{nguyen2019short} & General & Privacy/Security & Keywords & SVM & Yes \\
        Ebrahimi et al. \cite{ebrahimi2022unsupervised} & Finance & Privacy/Security & Keywords & - & - \\
        Nema et al. \cite{nema2022analyzing} & General & Privacy & Regex & BERT & Yes \\
        Harkous et al. \cite{harkous2022hark} & General & Privacy & Generic NLI & T5 & Yes \\
        \rowcolor{green!15} CMER (our approach) & Finance & Privacy/Security & Domain-specific NLI & Llama-like LLMs & No \\
        \hline
    \end{tabular}
\end{table*}

\noindent \textbf{Privacy and security in app reviews. }
The prior studies closest to ours are presented in Table \ref{tab:rw}. Besmer et al. \cite{besmer2020investigating} trained a logistic regression model to detect PSRs and showed that PSRs have lower ratings and more negative sentiments but higher engagement. Mukherjee et al. \cite{mukherjee2020empirical} and Nguyen et al. \cite{nguyen2019short} developed an SVM classifier to extract reviews related to privacy and security. 
Ebrahimi et al. \cite{ebrahimi2022unsupervised} conducted a manual analysis of MH app reviews using privacy-indicative keywords, revealing that privacy concerns vary by domain and are expressed with diverse language. Nema et al. \cite{nema2022analyzing} utilized regex for data sampling and developed a BERT model to classify privacy reviews on a large scale. They identified 440K privacy reviews from 2M apps across 29 app domains on the Google Play store.

In summary, most of the existing work employs traditional NLP approaches to develop classifiers and relies on keyword- or regex-based methods to sample the training data. These choices are intercorrelated, as it is easy to achieve high performance using traditional models on test sets created with such sampling methods \cite{harkous2022hark}. However, their reliance on pre-defined, context-independent keywords may not reflect the actual terminology used by reviewers and potentially lead to inaccuracies \cite{alomar2021finding}.
Harkous et al. \cite{harkous2022hark} utilized the NLI method to alleviate the limitations of keyword-based search techniques, conducting a detailed analysis of users’ privacy concerns in app reviews through established privacy taxonomies \cite{solove2005taxonomy, wang2009privacy}. They developed 31 privacy hypotheses for the NLI task, yet their approach predominantly employed generic privacy concepts, which can fail to identify finance domain-specific privacy and security concerns as users' ethical concerns are domain-dependent \cite{ebrahimi2022unsupervised}. Furthermore, all these studies rely on labeled datasets for fine-tuning classification models, highlighting a limitation in scenarios with limited or no labeled data.

\textbf{Our approach} differs from these studies in several ways. First, we focus specifically on privacy and security in the finance app domain. Second, we use NLI to sample potential PSRs instead of relying on a keyword or regex approach used by others \cite{besmer2020investigating, mukherjee2020empirical, nguyen2019short, ebrahimi2022unsupervised, nema2022analyzing}. Furthermore, we utilize finance domain-specific hypotheses for NLI instead of relying on the generic privacy hypotheses used in \cite{harkous2022hark}. Third, we leverage Llama-like LLMs to classify PSRs instead of relying on supervised classification methods \cite{nguyen2019short, nema2022analyzing, harkous2022hark}. Additionally, our approach does not require fine-tuning on the labeled dataset, thus reducing manual data labeling efforts. Finally, we highlight that our goal is to extract PSRs to determine privacy and security requirements and contribute to the field of RE, whereas other studies do not specifically focus on any RE-related tasks.

\noindent \textbf{LLMs}: 
Decoder-only LLMs such as GPT \cite{achiam2023gpt} and LLaMA \cite{touvron2023llama} have demonstrated the power and versatility of the transformer architecture when scaling up the number of parameters \cite{kaplan2020scaling}. In particular, they exhibit emergent abilities that arise suddenly at large scales and cannot be extrapolated from smaller models. After pre-training, many of these models undergo additional training with the RLHF technique \cite{ouyang2022training} to align with human objectives through reward feedback \cite{christiano2017deep}. This additional fine-tuning makes them a better choice for many NLP tasks where they need to follow user instructions \cite{ouyang2022training}. Furthermore, these models often need little to no fine-tuning on labeled data while generating relevant outputs. \cite{hou2023large}.

In the RE domain, decoder-only LLMs, have been widely used for various tasks, such as requirements classification \cite{arora2024advancing}, generation of new requirements \cite{bencheikh2023exploring}, requirements elicitation \cite{ronanki2023investigating, white2024chatgpt}, requirements analysis \cite{rasheed2024autonomous}, and resolving ambiguity in requirements \cite{sridhara2023chatgpt}. They have shown benefits with enhanced productivity and cost savings for various RE tasks \cite{marques2024using}. Despite such promising results for various RE tasks, Llama-like LLMs have not yet been leveraged in the domain of RE for the classification of PSRs, which we address in our study.

\section{Conclusion and Future Work} \label{conclude}
We present CMER (A \underline{C}ontext-Aware Approach for \underline{M}ining \underline{E}thical Concern-related App \underline{R}eviews), which enables developers to proficiently discern ethical concerns associated with their applications and enhance them towards being more trustworthy and responsible by leveraging user feedback. At the core of CMER, we utilize domain-specific NLI (providing domain-specific context awareness) and a Llama-like LLM with zero-shot learning (eliminating the need for any fine-tuning) to extract ethical concerns-related app reviews. We showed the efficacy of CMER by mining 2,178 privacy and security-related reviews from the mobile investment apps. Our findings also highlight that CMER can be adapted to other domains and ethical concerns, such as safety and transparency, with modifications to hypotheses. We believe that CMER advances the state-of-the-art towards a critical step in the requirements elicitation process by leveraging user feedback in app reviews.

In future work, we will (i) leverage topic modeling to automatically identify the main topics addressed by users in concern-related reviews; (ii) create a user-friendly and interactive tool for developers to extract concern-related reviews and summarize them easily; (iii) automatically extract requirements from the concern-related reviews which can be directly addressed and implemented in the development phase; (iv) devise an interactive guide in which practitioners can explore concern-related topics and navigate through relevant reviews to understand the evidence for each recommendation.

\section{Acknowledgement}
This research is supported and funded by the NSERC Alliance-Alberta Innovates Advance Program Stream I program. Also, we thank the anonymous reviewers for their constructive feedback, which has helped improve this work further.

\bibliographystyle{IEEEtran}
\bibliography{aire_ref}

\end{document}